# GRAVIMETRIC MEASUREMENT OF MAGNETIC FIELD GRADIENT SPATIAL DISTRIBUTION

Arutunian S.G., Dobrovolski N.M., Egiazarian S.L,
Mailian M.R., Sinenko I.G., Sinjavski A.V. Vasiniuk I.E.,
*Yerevan Physics Institute*
375036, Br. Alikhanian St. 2, Yerevan, Armenia
e-mail: femto@uniphi.yerphi.am

Magnetic interaction between a weighing sample and an external magnetic field allows to measure characteristics of magnetic field (a sample with known magnetic characteristics), as well as the magnetic properties of a sample (a known magnetic field). Measurement of materials' magnetic permeability is a well known application of this method [1, 2].

In this paper we restrict ourselves to the measurement of magnetic field spatial distribution, which was achieved by scanning of samples from known materials along the vertical axis. Field measurements by Hall's detector were done to calibrate obtained data. Such measurements are of great interest in some branches of physics, in particular, in accelerator physics, where the quality of magnetic system parts eventually determine the quality of accelerated bunches. Development of a simple and cheep device for measurement of magnetic field spatial distribution is an urgent problem [3]. The developed system for gravimetric measurement of magnetic field gradients partially solves this problem.

## 1. PRELIMINARY NOTES

The force of interaction $\vec{F}$ between a small sample of magnetic moment $\vec{M}$ with an external magnetic field of strength $\vec{H}(\vec{r})$ is defined by expression

$$\vec{F} = grad(\vec{M}\vec{H}). \tag{1}$$

We are mainly interested in two cases: when the magnetic moment is entirely induced and is determined by the magnetic permeability $\mu$ of the sample's material and, the second, when the magnetic moment is independent of the external field.

In the first case one has to take into account the proportionality of magnetic moment to the internal field $\vec{H}_i$ of the sample, which in its turn, is determined by the external magnetic field, magnetic susceptibilty and sample's geometric form. For an axially symmetric sample located in an axially symmetric magnetic field the interaction force is defined as follows [1]:

$$F_z = \frac{\mu - 1}{8\pi(1 + D(\mu - 1))} V \frac{\partial H_z^2}{\partial z}, \tag{2}$$

where $D$ is demagnetization factor determined by the sample's geometry. In so doing, we suggest that $\mu$ is independent of the magnetic field.

The force of interaction of a permanent magnet of magnetic moment $M_z$ with an external field $H_z$ is determined by the expression:

$$F_z = M_z \frac{\partial H_z}{\partial z}. \tag{3}$$

When studying samarium-cobalt magnets in the fields of the order of Earth's magnetic field one also can ignore the induced magnetization, i.e. one can consider $M_z$ as independent of the external field.

The scanning of a sample along the $z$ axis gives an information about the spatial distribution of the interaction force, which restores the gradient of the magnetic field square (2) or immediately the field's gradient up to a constant factor (3).

## 2. EXPERIMENTAL SETUP

### *2.1. String Tension Pickup*

A specially developed String Tension Pickup (STP) was used to measure the force of interaction between the sample and magnetic field. String magnetometer [2] with some improvements was taken as a prototype. In particular, by special means the lower end of the string with load was fixed in horizontal plane. As strings were used golden tungsten or beryl bronze wires of diameter 100 µm.

The magnetic system was made on the basis of samarium-cobalt permanent magnets of sizes 12×12×4mm and a magnetic circuit consisting of two symmetric parts from steel 3. As a result a C-form magnetic system was obtained, which closes the magnetic flow circuit through the gap with string. During the experiments pickups with 10 to 150 mm long strings were used.

An electromechanical generator excited the string oscillations due to the interaction of alternating current through the string with the magnetic field in the magnet gap. Positive feedback in the electric part of the generator excites and keeps self oscillations frequency close to its mechanical resonance.

The construction of STP allowed to tune normal oscillations frequency on each of first five harmonics. The first harmonic with loop in the middle turned to be the most convenient for measurements. The frequency corresponding to the first harmonic on the beryl bronze wire of diameter 100µm and length some 40mm was about 3kHz at the tension 3N. A system of forced automatic regulation keeps a stable current in the string, which allows to select its value to maximise the stability of its normal oscillations frequency. The frequency was multiplied by a factor 32 before measurement, and this along with normal oscillations high stability allowed to measure frequency with accuracy 0.03Hz during 1sec. Increasing the measurement time up to 10sec. allowed to achieve accuracy 0.003Hz.

Mentioned measures allowed to increase the level of relative sensitivity of STP up to $10^{-5}$ at the load ≤ 3N. Since the magnetometric measurements were done by slow rate scanning, it was necessary to provide long time stability at the level $10^{-5}$. A special method of fixation of the string's lower end practically excludes wire's drawing out of clips.

Pickup's thermostabilisation was done in two ways: either in analogue mode or by the programme using one of the eight digital outputs of the interface board. This allowed to achieve long-time stability during many hours. Fig. 1 shows a typical behaviour of pickup readings at permanent load 2.25N during more than 64 hours. During the mentioned measurements the pickup head was thermostabilised in ON/OFF mode by a programme via solid state relay using a signal from a thermistor. Since the thermistor was located on a massive base immediately near the heater, the temperature fluctuation was about 1 mV (2.5$^0$C), although the temperature of the base was stabilised with accuracy 0.1$^0$C.

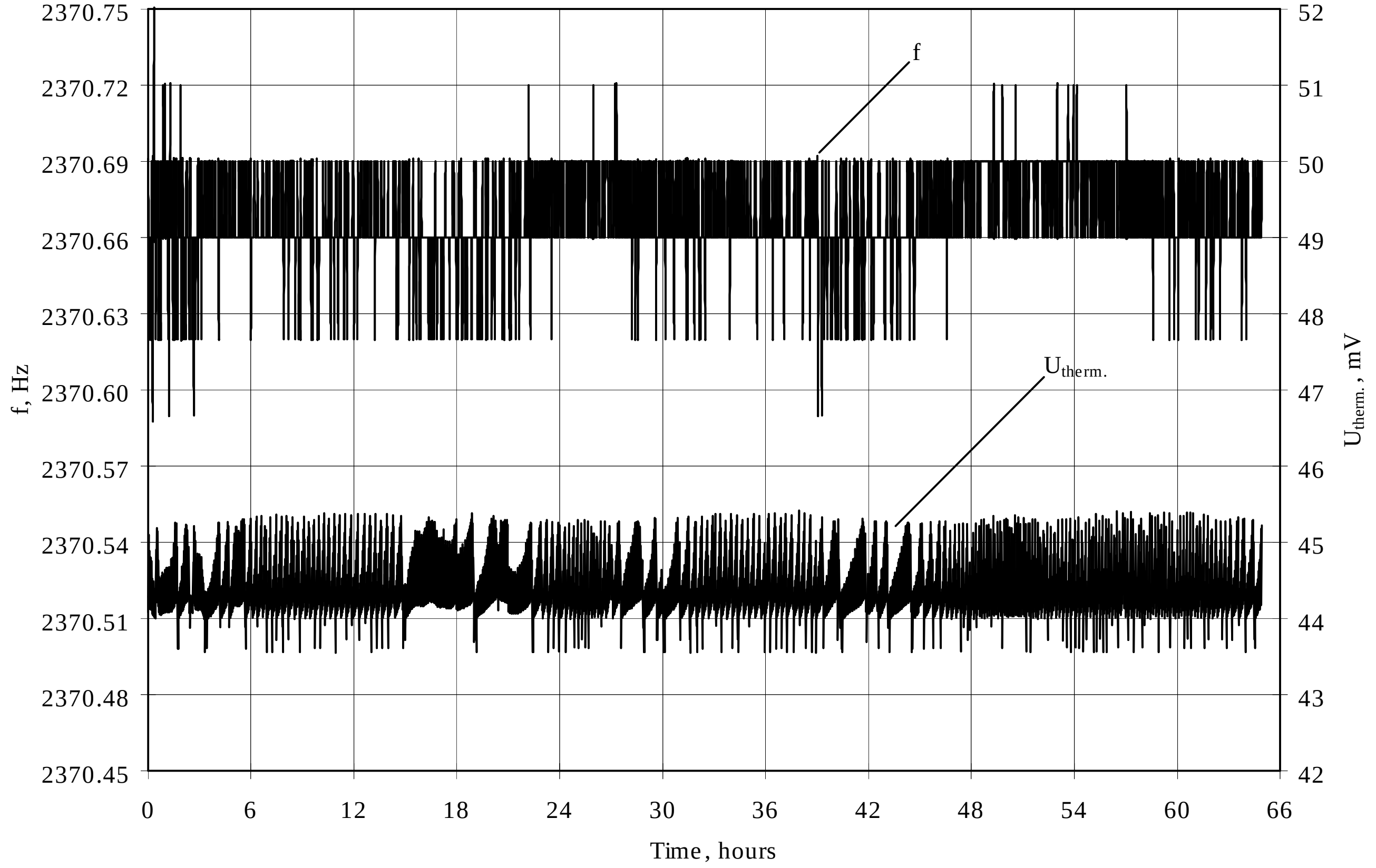

Fig. 1. Long time stability of the frequency of string strain gauge at constant load with pickup thermostabilisation. Voltage values of thermoregulation resistor in mV are also presented.

From Fig. 1 one can see that mean square deviation was $6.65\cdot10^{-6}$. This value is more promoted than that of known pickups based on measuring of vibrating wire frequency. E.g. in [4] string pressure pickups of resolution $10^{-4}$ from measurements interval are presented. String pressure pickups are used in oceanology investigations, and due to some improvement (pickup's thermostabilisation, string's preliminary ageing etc.) the error was lowered to the value $\pm1\div2\cdot10^{-4}$ ($\pm4\cdot10^{-5}$ is needed) [5]. The string pickups for measurement of tension in steel and concrete of the firm GeoKon have resolution $\pm3\cdot10^{-5}$ [6].

Thus, by means of mentioned construction of pickup we essentially improved the accuracy of measurements.

### *2.2. Scanner*

For scanning of magnetic field by moving the samples along $z$ axis a pulley-block lift system was made using a step engine $D\hat{S}I-200$ and low turn reductor. The rotation speed was regulated by PC. The speed of scanning varied from parts of mm to few mm per second.

### *2.3. Software*

The whole system was computerised. A computer board with two frequency inputs (up to 1.5MHz and 10MHz), three analogue inputs, eight digital outputs and a RS-232 port served as a base interface board. The system was tested on computers IBM 286 to Pentium. DOS and Windows programmes were developed. During the whole experiment the data from STP or Hall's detector were visualised on monitor, saved in file and processed later.

### *2.4. Samples*

Rings from soft magnetic steel, permalloy as well as of samarium-cobalt permanent magnets of sizes 12 × 12 × 4mm serve as samples. Note that usage of ferromagnetic samples in strong magnetic fields entails some difficulties caused by nonhomogenous and non-linear dependence between magnetisation of samples' material and the external field. Since the measurements were done in weak magnetic fields and had demonstration character, we had ignored this nonhomogenousity. To measure gradient of string magnetic fields one have to use para- or diamagnetic materials and the sensitivity of pickup allows to do it.

### *2.5. Hall detectors system*

To calibrate the gravimetric measurement system it was supplied by Hall's detectors. Note that to measure the weak magnetic fields by Hall's detector serious difficulties were overcame to eliminate of strays arising, in particular, from switching of current and potentials at each measurement.

Measurement of magnetic field by Hall detectors were done at alternating sine form current of frequency 1kHz (amplitude of the current through the pickup was about 100mA). Output signal was applied to the input of selective nanovoltmeter SN237 with narrow bandwidth and current gain $10^3$. Direct voltage signal from selective nanovoltmeter output was converted to match the interface board input. Polling of Hall's detectors was done during a few ms at each second. The current in coils was switched on/off via interface board regulation channel by means of a solid state relay, the "on" state time being 20sec., "off" - 40sec. This allowed to compare signals from Hall's detectors at on/off state of current in coils, in such way excluding the zero drift.

## 3. EXPERIMENTAL RESULTS

Experiments were done on two types of magnets: composite solenoid with symmetry axis along the $z$ axis and Helmholtz coils system designed to compensate Earth's magnetic field (coils lay in horizontal plane). Sizes of composite solenoid (two conjugated coaxial coils) were: inner diameter of coil windings was 25mm, external diameter - 31.2mm, length of each winding - 29mm, gap between them - 4mm, number of wraps in each winding - 1121. Helmholtz coils sizes were: diameter of each thin wrap was 900mm, distance between them - 460mm.

Each gravimetric measurement was a representation of interaction force between the sample and magnetic field by scanning of $z$ axis. Currents in solenoids and coils were selected to have such a value that the measured values lay in operating range of STP with optimal range of sensitivity. In case of composite solenoid this current was $I$=200mA, for Helmholtz coils $I$=3.1A. Since both systems consist of two separated coils, the measurements were done for parallel and antiparallel currents. Fig. 2 represents the primary experimental results for composite solenoid for parallel (curve A) and antiparallel (B) currents respectively. Vertical axis represents the interaction force in mN. The scanning speed was 0.113mm/sec. A steel-3 ring with inner diameter 3mm, external diameter 13mm and thickness 6.4mm was used as a sample. Signals at going in and out of the solenoid were coincide in satisfactory level (null signals with switched off currents are omitted). The width of experimental track is about 0.1mN and is greater than the resolution of string tension pickup and is mainly caused by swinging of rather big sample in the narrow hole of coils during the scanning. Division of $F$ by the factor $(\mu - 1)V/8\pi(1 + D(\mu - 1))$ gives dependence of the value $\partial H_z^2/\partial z$ on $z$. In its turn the last function allows to restore the function $H_z(z)$ along the axis of solenoid.

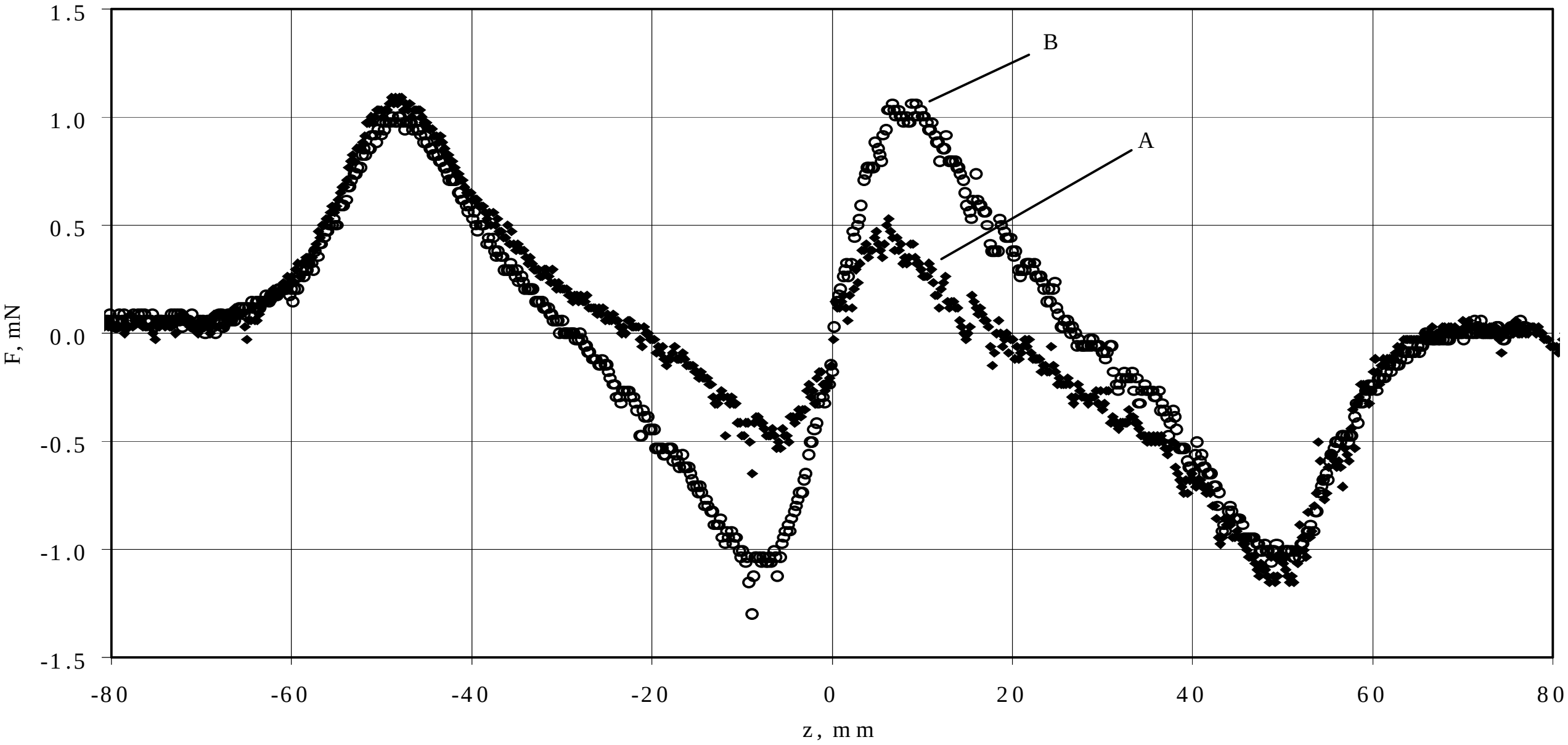

Fig. 2. Gravimetric measurements for composite solenoid (A - parallel currents, B - antiparallel currents). Since the interaction with magnetic field is determined by the square of field gradient, the spatial distribution of measurements for parallel and antiparallel currents have similar behaviour.

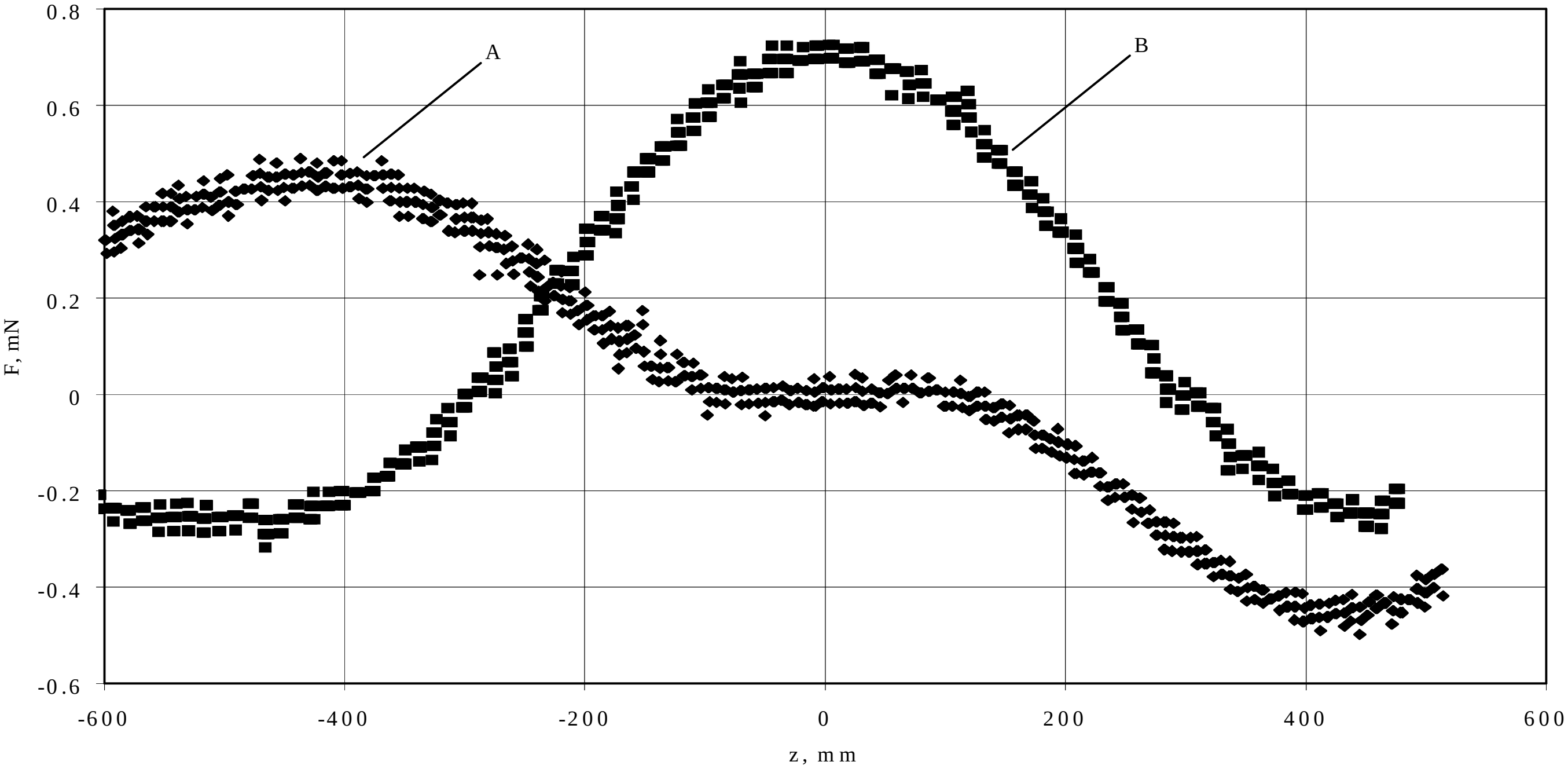

Fig. 3 Gravimetric measurements for Helmholtz coils system
(A - parallel currents, B - antiparallel currents).

Fig. 3 represents similar gravimetric measurements for Helmholtz coils system (A corresponds to parallel currents, B - to antiparallel ones). Samarium-cobalt magnets were used as samples. Scanning speed was 0.394mm/sec. Here the track width was about 0.04mN. This is less than that in previous case, since the samples relative sizes were much more less than coils diameter. Presented curves define the Helmholtz coils system magnetic field gradient up to a constant.

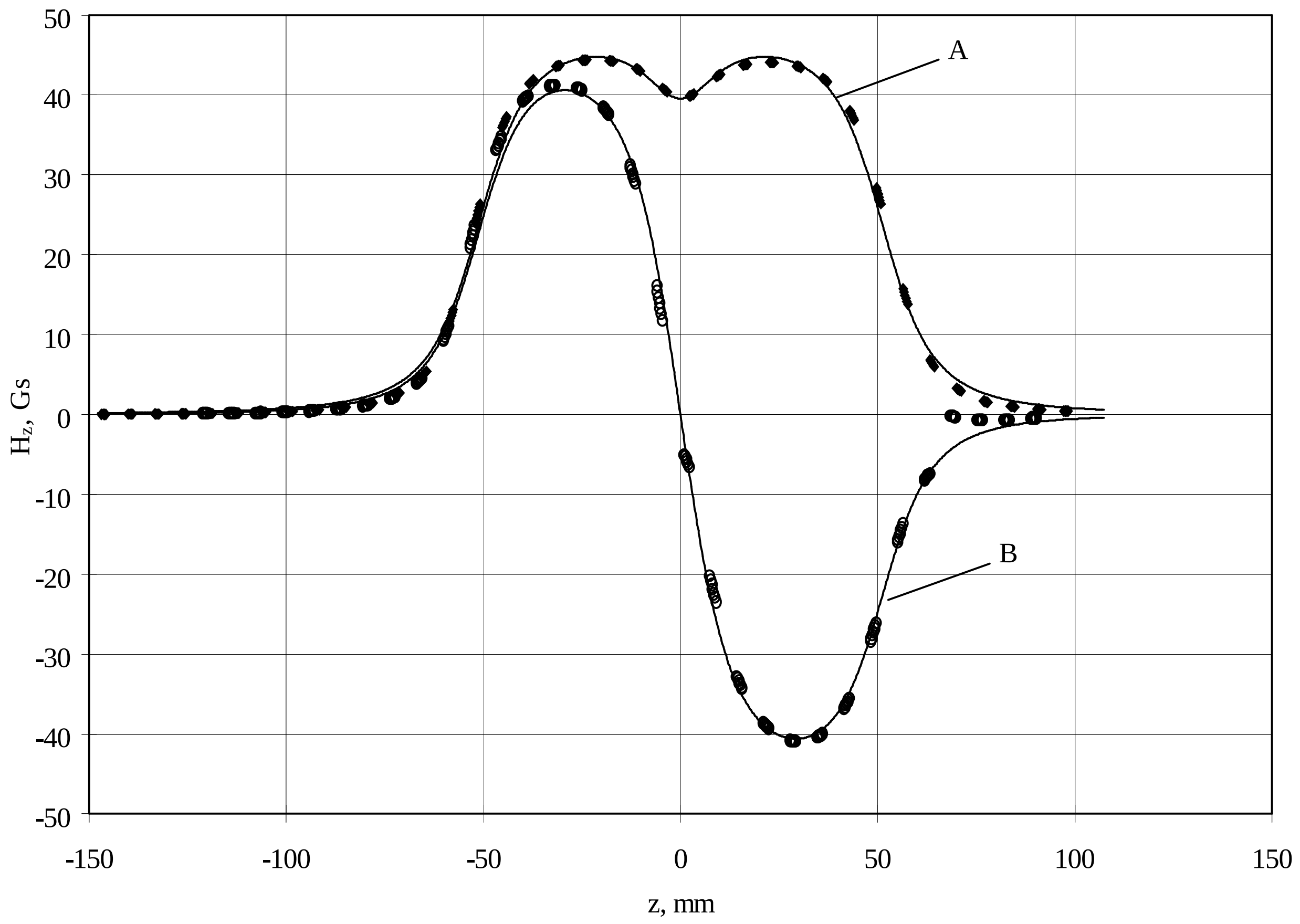

Fig. 4. Hall measurements for composite solenoid
(A - parallel currents, B - antiparallel currents).

Hall measurements of magnetic field for composite solenoid are shown in Fig. 4 (A - parallel currents, B - antiparallel). Experimental data as well as calculated curves are presented (solenoid parameters were refined by adjusting of experimental points). There were some difficulties with measurement of magnetic field sign, because the selective nanovoltmeter measures the amplitude of alternating output signal.

Similar Hall measurements of Helmholtz coils system are presented in Fig. 5.

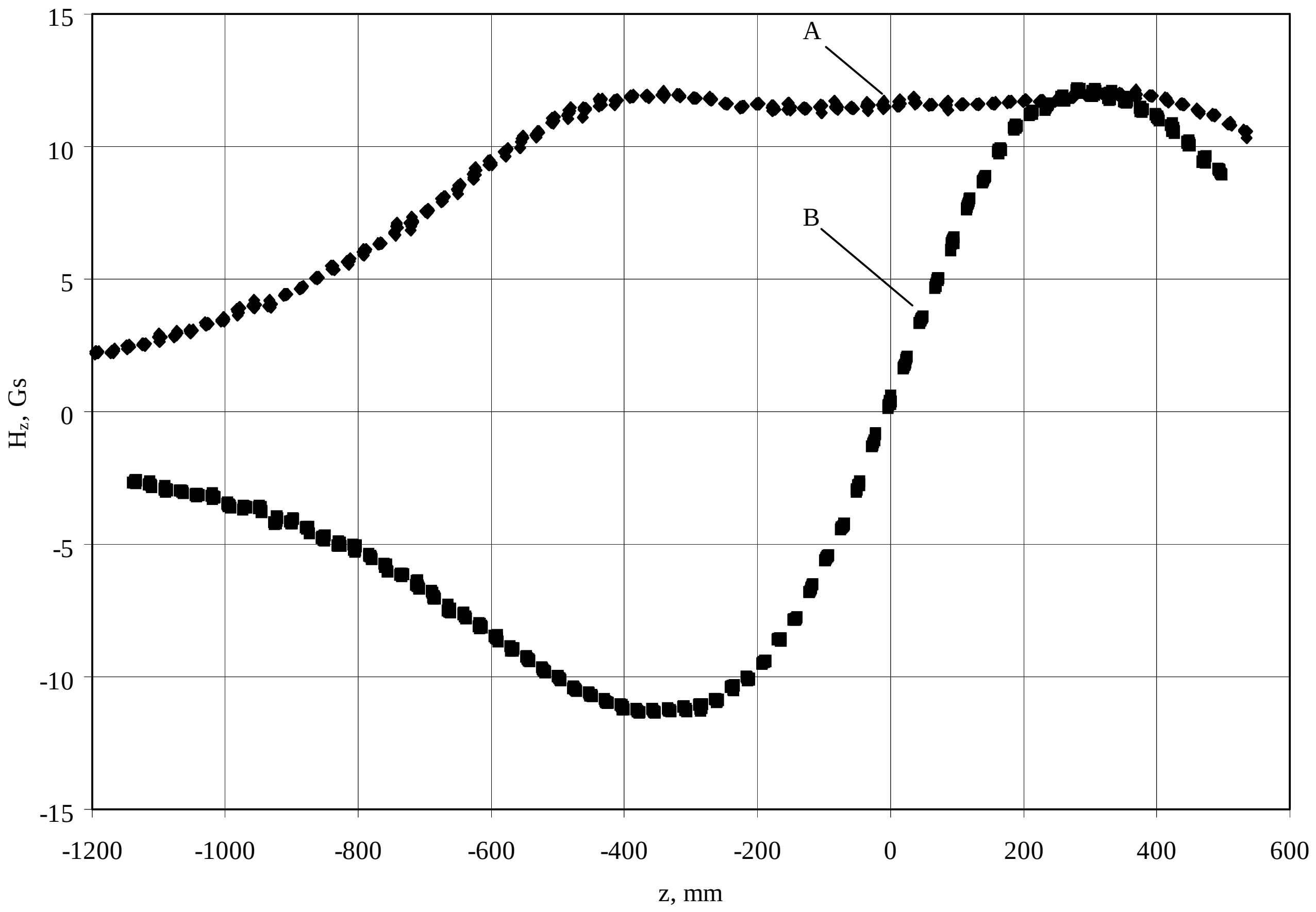

Fig. 5. Hall measurements for (A - parallel currents, B - antiparallel currents).

## 4. PROCESSING OF EXPERIMENTAL RESULTS

Obtained experimental results were processed to define the magnetic field gradients. To find it actually one have to find the factor $(\mu-1)V/8\pi(1+D(\mu-1))$ for samples with magnetic susceptibility $\mu$ or magnetic moment $M_z$ for samples from permanent magnet. In principle, one can estimate these parameters using the tabulated values of known materials in use, however, since these factors depend on the samples shape, finding of these parameters using special calibrating measurements by Hall detectors seemed preferable. Note that measurements done near the experimental points of field gradient are sufficient.

There are two way to compare the gravimetric measurements with Hall's detectors ones: integrate gravimetric curves or differentiate Hall detectors ones. Taking into account that during each period of measurement the current trough coils was switched off, the numerical differentiation is preferable, because it uses information of local sections of experimental curves.

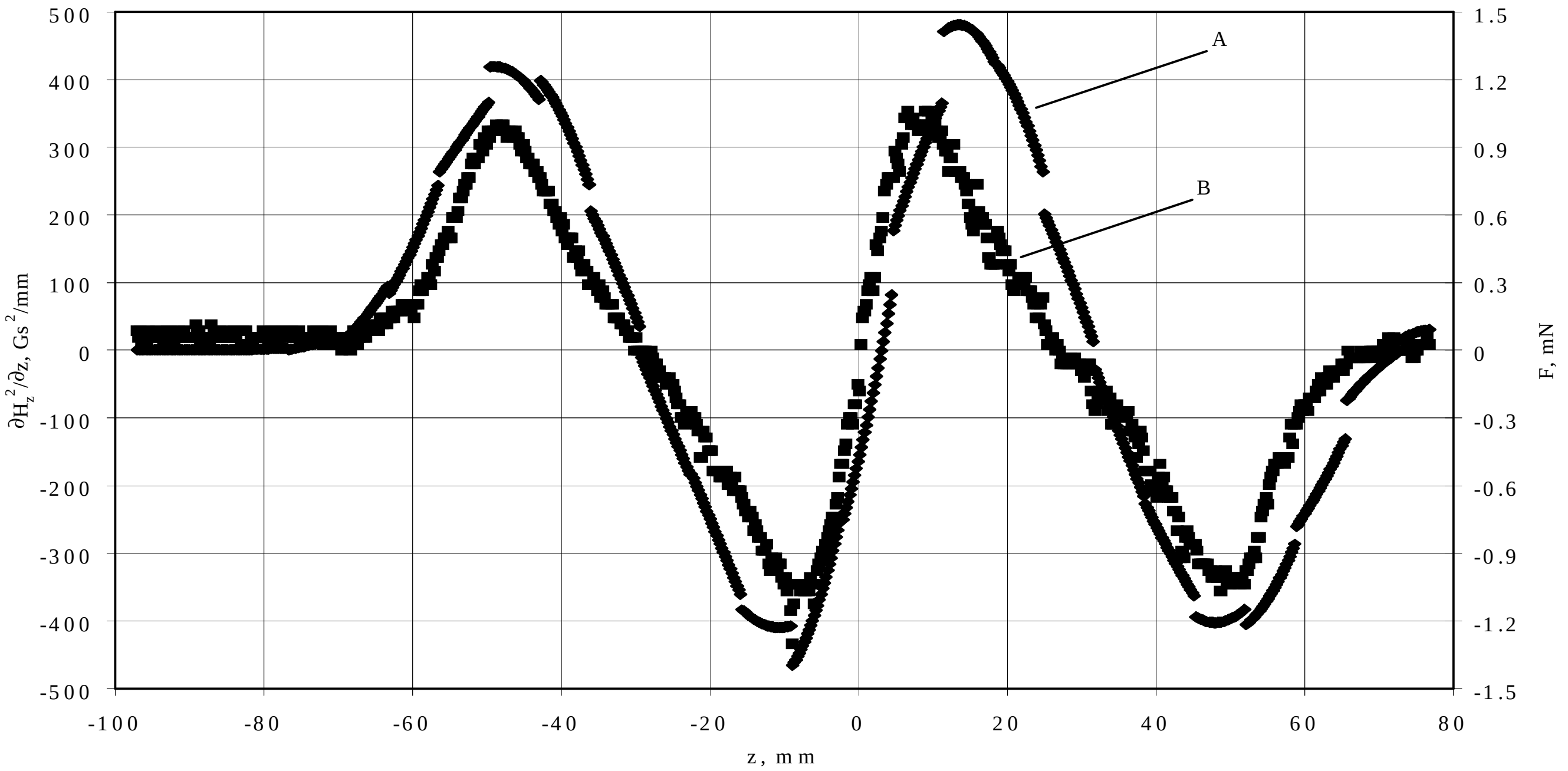

Fig. 6. Gradient of field square for composite solenoid restored by Hall measurements (A). Comparison with gravimetric measurements (B).

To calibrate measurements of composite solenoid the experimental results for antiparallel currents were used. Calculation of gradient of magnetic field square using experimental points for composite solenoid was done by numerical approximation of seven groups of currents switching on. Results were compared with gravimetric measurements and are presented in Fig. 6. Such a comparison gives a coefficient of proportionality between gradient of magnetic field square and interaction force equal to 445.48 ($Gs^2$/cm)/mN with correlation factor 87%.

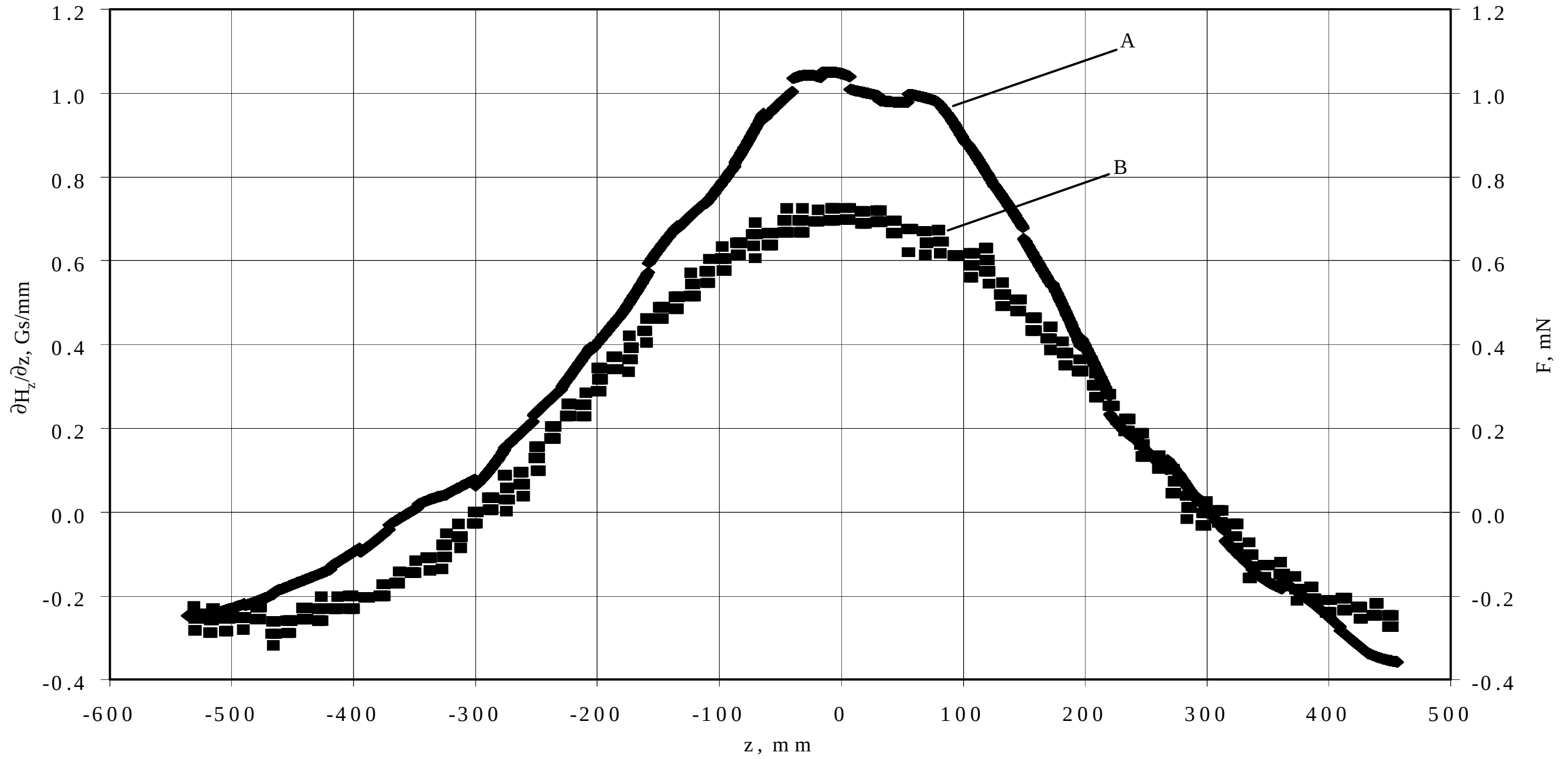

Fig. 7. Gradient of field for Helmholtz coils system restored by Hall measurements (A). Comparison with gravimetric measurements (B).

Similar calculations of the field gradient for the Helmholtz coils system using Hall's detector measurements are presented in Fig. 7. In this case comparison with the gravimetric measurements defines the samples magnetic moment of the order of 1.35 (Gs/cm)/mN with correlation factor 95%.

In further experiments with the same samples Hall measurements obviously are not needed.

## 5. CONCLUSION

This work was aimed to develop and construct a complete system of measurements, although the measurements were done on demonstration samples. Varying of temperature both to the high and low temperatures are possible and will essentially broaden unit's possibilities.

Combination of these measurements with other methods of measurement of magnetic field characteristics gives a possibility to fulfil sufficiently simple and precise measurements of magnetic parameters.

This work was fulfilled by support of firm HTM Reetz (Berlin), and authors are thankful to the employees of the firm and especially to Dr. R.Reetz. Authors also thank V.Gavalian and A.Aleksanian for their help.

## REFERENCES

1. Methods of Experimental Physics, v.1, Classical Methods, - Ed. by I.Estormann, Acad.Press, 1959, N7, p. 537
2. El'tsev Ju. F., Zakosarenko V.M., Tsebro V.I., String Magnetometer, Trudy FIAN, v. 150, M, 1984.
3. A. Harvey, Proc. Magnet measurement and alignment CERN Accelerator School, Ed. S. Turner (Switzerland, 16-20 March 1992), CERN 92-05, p. 228-239.
4. Asch G. Les Capteurs en Instrumentation Industrielle, v. 1, 2. Dunod, 1991.
5. Kovchin I.S. Autonomous Oceanographic means of Measurement.. - L, Gidrometeoizdat, 1991.
6. Geokon incorporated.- http://www.geokon.com/products.